\documentclass[12pt]{iopart}

\usepackage{iopams}
\usepackage{graphics,wrapfig,times}
\usepackage{graphicx}

\usepackage{epsfig,epstopdf}

\usepackage{amsfonts}
\usepackage[T1]{fontenc}

\def\bar{\overline}
\def\be{\begin{equation}}
\def\ee{\end{equation}}
\def\bea{\begin{eqnarray}}
\def\eea{\end{eqnarray}}
\def\ba{\begin{array}}
\def\ea{\end{array}}
\def\a{\alpha}
\def\b{\beta}
\def\nn{\nonumber}
\def\c{\gamma}
\def\d{\delta}
\def\w{\wedge}
\def\up{\stackrel}
\def\dd{\textrm{d}}

\def\td{\textrm{d}}
\def\te{\textbf{e}}
\def\tb{\textbf{b}}
\def\half{\frac{1}{2}}
\def\cc{\tilde{c}}
\def\cs{c}

\begin{document}

\title[Quasi-local mass in the Newtonian space-time]{Quasi-local mass in the covariant Newtonian space-time}

\author{Yu-Huei Wu$^1$ and Chih-Hung Wang$^2$}

\address{$^1$Institute of Astronomy, National Central University,
     Chungli, Taiwan 320, R. O. C.}

\address{$^2$Department of Physics, National Central University,
     Chungli, Taiwan 320, R. O. C.}
\ead{$^2$\mailto{chwang@phy.ncu.edu.tw},
$^1$\mailto{yhwu@astro.ncu.edu.tw}}

\begin{abstract}

In general relativity, quasi-local energy-momentum expressions
have been constructed from various formulae. However, Newtonian
theory of gravity gives a well known and an unique quasi-local
mass expression (surface integration). Since geometrical
formulation of Newtonian gravity has been established in the
covariant Newtonian space-time, it provides a \textit{covariant}
approximation from relativistic to Newtonian theories. By using
this approximation, we calculate Komar integral, Brown-York
quasi-local energy and Dougan-Mason quasi-local mass in the
covariant Newtonian space-time. It turns out that Komar integral
naturally gives the Newtonian quasi-local mass expression,
however, further conditions (spherical symmetry) need to be made
for Brown-York and Dougan-Mason expressions.

\pacs{04.20.Cv, 04.20.Gz, 04.25.Nx }
\end{abstract}
 \submitto{\CQG}

\maketitle

\section{Introduction}

General Relativity (GR) is a theory for describing the dynamics of
metric $g$ and all other classical matter fields. Unlike energy
and momentum densities of matter fields which are well-defined,
gravitational energy cannot be localized because of the
equivalence principle. Quasi-local idea is to define gravitational
energy-momentum associated to a closed 2-surface (see the review
article \cite{Szabados}). Relativistic expressions of quasi-local
mass, energy, and angular momentum have been formulated from
various approaches. However, Newtonian theory of gravity exists an
unique and physically well-defined expression of quasi-local mass.
Since GR will yield Newtonian theory of gravity in a certain
limit, \textit{it is reasonable to believe that all of quasi-local
expressions should return to the Newtonian quasi-local mass
expression in the same limit.} Newtonian and post-Newtonian
approximations are asymptotic to general relativity by using the
Newtonian sequence method \cite{Futamase&Schutz}, and hence Jeffryes
\cite{Jeffryes} use it to verify the Newtonian limit of Penrose quasi-local
mass. Unfortunately, it can return to the Newtonian mass and
energy only by using transaction between 2-surface twistors and
3-surface twistors. Dougan-Mason mass \cite{Dougan&Mason} is
another expression based on spinors and it has weak-field limit by
transferring 2-surface integral into 3-volume integral. It is not
clear to us that these quasi-local expressions with associated
spin frames can \textit{directly} return to the Newtonian
quasi-local mass expression in the Newtonian space-time.

The post-Newtonian theory provided a scheme to obtain an
approximate solution of Einstein field equations and equations of
motion based on two limits: slow motions and weakly gravitational
field. Though Newtonian dynamics can be recovered in this scheme,
it is not a covariant approach since coordinate condition
(harmonic coordinates) and background Minkowski metric are
required. However,  Carter \textit{el al.} \cite{Carter}
established covariant Newtonian multi-fluids and (magneto)-elastic
solid models for neutron star crust by using geometric formulation
of Newtonian gravity which has been largely developed by Kunzle
\cite{Kunzle}. Furthermore, if one try to probe the Neutron star,
this requires a more compatible treatment between relativistic and
Newtonian theories since the inner core may need to be described 
in terms of GR and it is sufficient to use the Newtonian theory for the outer crust. The
covariant Newtonian space-time $\mathcal{N}$ is considered as a
\textit{flat} Euclidean 3-space $\Sigma$ direct product a
one-dimensional time manifold $t$ and characterized by a
degenerated (contravariant) flat-metric $\gamma$ and a one-form
$\dd t$. It should be stressed that the metric-compatible and
torsion-free connection $\nabla$ cannot be uniquely determined
\cite{Kunzle} by $\gamma$ and $\dd t$, so it is not necessary to
yield a flat connection, i.e., $\nabla_{X_a} \nabla_{X_b} =
\nabla_{X_b} \nabla_{X_a}$. One can introduce the Newton-Cartan
connection $D$, instead of the flat connection, to obtain a
geometrical formulation of Newtonian theory \cite{Carter}. The
detail construction will be reviewed in the following section.

Our purpose in this paper is to examine Komar integral, Brown-York
(BY) expression, and Dougan-Mason (DM) mass in the covariant
Newtonian space-time. In \cite{Carter}, it yields a
four-dimensional metric $\tilde{g}=- (c^2 + 2 \Phi) \,\dd t
\otimes \dd t + \eta$ in a pseudo-Riemannian spacetime
$\mathcal{M}$ and when speed of light $c$ approaches to the limit
$c \rightarrow \infty$, one obtains $\{ \tilde{g}/c^2,
\,\tilde{g}^{-1}, \,\check{{\nabla}}\} \rightarrow \{ \dd t,
\,\gamma,\, D\}$. Here, $\Phi$ is considered as the Newtonian
potential and $\eta$ denotes the (covariant) flat-metric on
$\Sigma$. This provides a covariant transition between
relativistic and covariant Newtonian theories. We first use the
metric $\tilde{g}$ to calculate the three different quasi-local
expressions and it turns out that Komar integral can exactly yield
the Newtonian quasi-local expression, but BY and DM's energies
expressions give zero. The result implies that the measure of
these two quasi-local energy requires the information of non-flat
metric on $\Sigma$ and leads us to ask a question:\textit{ can a
\textit{purely} Newtonian type of star distort the geometry of
$\Sigma$, i.e., generate pseudo-Riemannian curvature on it?} It has
been mentioned that $\tilde{g}$ is sufficient to represent the
relativistic corresponding metric of the covariant Newtonian
theory. However, the post-Newtonian theory indicated that a
Newtonian star should generate a spatial non-flat 3-metric in the
leading order $\Or(1/c^2)$. Thus, the Newtonian limit of GR will
actually yield a non-flat 3-dimensional hypersurface. It leads us
to generalize $\tilde{g}$ by including a leading-order correction
(a covariant 3-metric $h$) in $\eta$ which gives $\hat{g}=- (c^2 +
2 \Phi) \,\dd t \otimes \dd t + (\eta + h/c^2)$. It is easy to
show that $\hat{g}$ can also return to the covariant Newtonian
space-time $\{ \dd t, \,\gamma,\, D\}$ in the limit $c \rightarrow
\infty$. The three quasi-local expressions will be calculated by
using $\hat{g}$ instead of $\tilde{g}$.

The plan of this paper is as follows. In \Sref{Milne}, we give a
review of covariant Newtonian space-time and Newton-Cartan
connection which is associated to the Milne group. In \Sref{RelC},
we consider a specific form of non-flat 3-metric $h$ which consists
of two unknown functions in the 2+1 decomposition and are determined
by Newtonian Poisson equations. The solution agrees with
weak-field of Schwarzschild solution when one imposes spherical
symmetry on it. In \Sref{Quasi}, we present a detail calculation
of Komar, BY and DM's quasi-local energy by using $\hat{g}$ and also
find the asymptotic Newtonian spin frame. It should be notice that the convention is chosen $(+---)$ in \Sref{DM} due to the spinor algebra
usage.


\section{Milne Structure of Newtonian Space-time} \label{Milne}

Since Einstein developed the special theory of relativity,
Euclidean 3-space and time are merged into a four-dimensional
manifold called Minkowski space-time which is characterized by a
flat metric $g^M$ and Levi-Civita connection $\check{\nabla}$. Later on, general
relativity (a relativistic theory of gravity) was then generalized
to four-dimensional curved spacetime. Though general relativity
gives the Newtonian limit at a slow-motion approximation, it is a
coordinate dependent approach, i.e., it requires to introduce the
specific coordinates. To amend this unsatisfaction, the covariant
description of Newtonian space-time provides a more consistent
transition between relativistic and Newtonian theory of gravity.
Furthermore, some problems of astrophysics require both
relativistic and Newtonian treatments, e.g. inner core and outer
curst of a neutron star, so to facilitate the transition can also
improve solving these problems.

Newtonian (Galilean) space-time is considered as a direct product of
a flat Euclidean 3-space $\Sigma$ and one dimensional Euclidean
time line $t$, i.e., four dimensional fiber bundle. Each fiber
$\Sigma_t$ contains a Euclidean (contravariant) flat metric
$\gamma$ and one-form $\dd t$ where $\gamma$ is a degenerate metric
and cannot upper or lower indices, i.e. no metric-dual definition
of tensor fields. It has been found that conditions $\nabla
\gamma=0$ and $\nabla \td t=0$ with vanishing torsion cannot give
an unique connection $\nabla$ in the Newtonian space-time
\cite{Kunzle}. Nevertheless, $\nabla$ can be uniquely determined
by introducing an ether frame $\te$ which is a unit timelike
vector field $\dd t(\te)=1$. By using
\be\ba{lll} \nabla_\rho \c^{\mu\nu} & =&  \partial_\rho  \c^{\mu\nu} + 2 \Gamma^{(\mu}_{\rho\a} \c^{\nu)\a}=0, \\
\nabla_\a t_\b &=& \partial_\a t_\b - \Gamma^\rho_{\a\b} t_\rho=0,
\label{metric_compatible}\ea\ee
the torsion-free connection components $\Gamma^\rho_{(\a\b)}$ can be obtained as
\be
  \Gamma^\c{_{\a\b}} = \c^{\c\rho} \partial_\rho e^\sigma \eta_{\sigma(\a} t_{\b)}
 - \eta_{\rho(\a} \partial_{\b)} \c^{\rho\c} + \half \c^{\c\rho} \partial_\rho \c^{\mu\nu}\eta_{\mu\a} \eta_{\nu\b}
  - t_{(\a}\partial_{\b)} e^{\c}\label{Kunzle-conn}
\ee where $\eta_{\a\b}$ are the components of covariant 3-metric $\eta$ satisfying
\be \eta(\textbf{e}, \,-)=0, \hspace{1.0cm}
\eta_{\mu\rho}\c^{\rho\nu} = \delta_\mu{^\nu} - e^\nu t_\mu. \ee
Greek indices ($\a, \b,\cdots=0,1,2,3$) denote the components of
tensor fields with respect to coordinate basis $\{\partial_\mu\}$
and co-basis $\{\td x^\nu\}$ and round brackets indicate index
symmetrization. Since $\gamma$ and $\dd t$ are flat metric, there
should exist an Aristotelian coordinate system $\{t, X^1,X^2,
X^3\}$ in which $\gamma^{\mu\nu}= \delta_1{^\mu} \delta_1{^\nu} +
\delta_2{^\mu} \delta_2{^\nu}+  \delta_3{^\mu} \delta_3{^\nu}$,
$t_{\mu}=\delta{_\mu}{^0}$, and also the corresponding ether frame
$e^\mu = \delta_0{^\mu}$ with $\eta_{\mu\nu}= \delta_\mu{^1}
\delta_\mu{^1} +  \delta_\mu{^2} \delta_\nu{^2}+  \delta_\mu{^3}
\delta_\nu{^3}$. \Eref{Kunzle-conn} then shows that
$\Gamma^\c{_{\a\b}}$ vanish in Aristotelian coordinates and thus
$\nabla$ is a \emph{flat} connection, i.e. $\nabla_{\partial_\mu}
\nabla_{\partial_\nu} - \nabla_{\partial_\nu}
\nabla_{\partial_\mu}=0$.

Besides coordinate transformations, there exist gauge
transformations corresponding to the freedom of choosing $\te$. It
is easy to verify that $\nabla$ is invariant under Galilean gauge
transformations
\be \te \rightarrow \bar{\te} = \te + \tb \label{Gauge_transformation}\ee
where $b^\mu$ are constants and satisfy $\td t (\tb)=0$. This
corresponds to the invariance of Newtonian mechanics under
Galilean transformations. To establish a covariant Newtonian
theory of gravity in terms of geometry, Carter \textit{et al.}
\cite{Carter} consider the Milne structure which is preserved by
time-dependent Galilean gauge transformations (Milne gauge
transformations), i.e. $b^\mu=b^\mu(t)$, in the Newtonian space-time.  It follows that $\gamma^{\mu\nu}\nabla_{\partial_\nu} \tb
=0$. In the Milne structure of Newtonian space-time,  $\nabla$ is
not gauge invariant and has the following gauge transformations
\be \Gamma^\mu{_{\nu\alpha}} \rightarrow
\bar{\Gamma}^\mu{_{\nu\alpha}}= \Gamma^\mu{_{\nu\alpha}} - a^\mu
t_\nu t_\alpha. \label{Connection_Transformation} \ee
Linear acceleration $\textbf{a}$ is defined by $\textbf{a}:=
\nabla_\te \tb $  and $a^\mu(t) =\td x^\mu(\textbf{a})$. It is
worth to stress that the Milne gauge-dependent connection is still
a flat connection.

In the Newtonian theory of gravity, gravitational field
$\textbf{g}$ can be described in terms of the Newtonian potential
$\Phi(x^\mu)$
\be \textbf{g}=- (\gamma^{\mu\nu} \nabla_{\partial_\nu}
\Phi)\partial_\mu \ee
and therefore $\dd t(\textbf{g})=0$. If one believes that linear
acceleration $\textbf{a}$ and $\textbf{g}$ cannot be distinguished
from any local Newtonian observer (a Newtonian type of equivalence
principle), it then follows that Newtonian mechanics should be
invariant under Milne gauge transformations. To obtain this
invariant property, it is necessary to introduce a Milne
gauge-invariant connection $D$ (Newton-Cartan connection)
\be \omega^\mu{_{\nu\alpha}}= \Gamma^\mu{_{\nu\alpha}}  - g^\mu
t_\nu t_\alpha. \label{NC_Connection}\ee
 and $\textbf{g}$ must have following gauge transformations
\be \textbf{g} \rightarrow \bar{\textbf{g}} = \textbf{g} -
\textbf{a}, \hspace{2.0cm} \Phi \rightarrow \bar{\Phi} = \Phi +
\alpha \ee
to compensate transformations \eref{Connection_Transformation},
where $a^\mu= \gamma^{\mu\c}\nabla_{\partial_{\c}} \alpha$.
According to \eref{metric_compatible} and \eref{NC_Connection},
one can verify that $D$ is metric-compatible  and torsion-free
connection $D \dd t = D \gamma=0$. We then further to calculate
the curvature tensor
\be \textbf{R}(\partial_\mu,
\partial_\nu) \partial_\alpha:= D_{\partial_\mu} D_{\partial_\nu}
\partial_\alpha - D_{\partial_\nu} D_{\partial_\mu}
\partial_\alpha
\ee which gives non-zero curvature 2-forms
\be R^\gamma{_{\mu}}= - t_\mu (\nabla_{\partial_\nu}
\textbf{g})^\gamma  \td x^\nu \wedge \td t \ee
and Ricci 1-forms
\be P_\alpha = i_{\partial_\mu}
R^\mu{_{\alpha}}=- t_\alpha (\nabla_{\partial_\mu}
\textbf{g})^\mu\, \td t, \label{Ricci_1-forms}
\ee
where $i_{\partial_\mu}$ denotes interior derivative. To complete
the geometric formulation of Newtonian gravity, one introduces the
field equation \be P_\alpha = 4 \pi G \rho^\mu t_\mu\,t_\alpha \td
t \ee which is equivalent to the Newtonian gravitational equation
\be \gamma^{\mu\nu} \nabla_{\partial_\mu} \nabla_{\partial_\nu}
\Phi = 4 \pi G \rho \label{Poisson_equation} \ee
where $G$ is the Newtonian gravitational constant. $\rho^\mu$
denotes mass flux 4-vector and mass density $\rho=\rho^\mu t_\mu$.
In terms of familiar vector calculus on $\Sigma$,
\eref{Poisson_equation} can be expressed as Poisson equation
$\vec{\nabla} \cdot \vec{\nabla} \Phi = 4 \pi G \rho$.

It is convenient to define a 3-dimensional Hodge map \# on
$\Sigma_t$ with $\textrm{\#} 1 = e^1 \wedge e^2 \wedge e^3$.
$\{e^i\}$ are orthonormal co-frames associated with 3-metric
$\eta$ and Latin indices $i, j, \cdots =1, 2, 3$. Without losing
any generality, one can simply consider $\rho=\rho(x^i)$ and then
the covariant expression of Newtonian quasi-local mass yields
\bea M &=& \int_{\Sigma_{t}} \rho^\a t_\a \,\td V \nn\\
& =&  \frac{1}{4\pi G} \int_{\Sigma_{t}}  \td \,\textrm{\#} \,\td
\Phi
= \frac{1}{4\pi G} \oint_{S=\partial \Sigma_{t}} \textrm{\#} \,\td \Phi \nn\\
 & =&  - \frac{1}{8\pi G} \oint_{S}  g^\mu\, n_\mu \,\sqrt{|\sigma|}\,\varepsilon_{a b}
 \,\td x^a \,\td x^b  = \frac{1}{4\pi G} \oint_S \vec{\nabla} \Phi \cdot {\textbf{n}} \,\td S \label{Newtonian_QLM}
\eea
where $x^a$ ($a, b =2,3$) are coordinates on the closed
2-surface $S$ and $\textbf{n}$ is a space-like $(\dd
t(\textbf{n})=0)$ unit outward normal to $S$. $\varepsilon_{a b}$
denotes Levi-Civita $\varepsilon$-symbol. It is worth to point out
that \eref{Newtonian_QLM} is valid for an arbitrary distribution
of mass density $\rho(x^i)$  on $\Sigma_t$ and Milne gauge
invariant.

\section{Relativistic Correspondence}\label{RelC}

It has been approved that most of relativistic theories of gravity
should regain the Newtonian theory of gravity in the limit $c
\rightarrow \infty$. In the post-Newtonian approximation, the
leading order of Einstein field equations yields Newtonian Poisson
equation in harmonic coordinates and it turns out that Newtonian
potential $\Phi$ appears in metric components $g_{00}= - c^2 - 2
\Phi$ and $g_{ij}=(1 - 2 \Phi / c^2)\delta_{ij}$.  Though $\Phi$
in $g_{00}$ is sufficient to represent the Newtonian limit, it
indicates that a Newtonian type of star will actually generate a
non-flat spatial metric. Since the covariant Newtonian theory has
been established, it provides a covariant approach from a
corresponding relativistic theory. Carter \textit{et al.}
\cite{Carter} considered the following four dimensional
non-degenerate metric
\be \tilde{g} = \eta - \tilde{c}^2 \,\td t \otimes \td t
\label{Rel-New-m}
\ee with its inverse $\tilde{g}^{-1} = \gamma - \tilde{c}^{-2}\,\,
\te \otimes \te$ in a pseudo-Riemannian manifold $\mathcal{M}$,
where $\tilde{c}^2= c^2 + 2 \Phi$.   It can be shown that $c
\rightarrow \infty$ gives
\be \tilde{g}^{-1} \rightarrow \gamma, \hspace{2.0cm}
{c}^{-2}\tilde{g} \rightarrow - \td t \otimes \td t,
\hspace{2.0cm} \check{\nabla} \rightarrow  D, \ee
where Levi-Civita connection $\check{\nabla}$ yields the
Newton-Cartan connection $D$. Though \eref{Rel-New-m} is
sufficient to represent the corresponding relativistic metric of
covariant Newtonian theory, it seems inconsistent with the
post-Newtonian approximation which gives a non-flat spatial
metric. Furthermore, BY and DM's quasi-local energies give zero
since $\gamma$ and $\eta$ are flat 3-metric. It leads us to
generalize $\tilde{g}$ by including leading order correction on
$\eta$. This generalization will obviously preserve the covariant
Newtonian theory in the limit $c \rightarrow \infty$.

Since the calculation of quasi-local quantities is associated with
a closed 2-surface $S$, it is useful to apply 2+1 decompositions
on 3-metric which gives
\be \eta= \tilde{n} \otimes \tilde{n} + \sigma, \hspace{1.0cm}
\gamma= n \otimes n + \tilde{\sigma} \ee with $\dd t (n)=\sigma(n,
-)=0$. $\tilde{n}=\tilde{g} (n, -)$ denotes the metric dual of the
space-like unit vector $n$ and $\sigma$ corresponds to the metric
on a family of closed 2-surface $S_n$ with their induced
contravariant metric $\tilde{\sigma}= \sigma^{\a\b} \partial_\a
\otimes \partial_\b$. Thus, the generalization of $\tilde{g}$ in
terms of 2+1 decompositions yields
\bea
\hat{g} &=& - \tilde{c}^2 \,\td t \otimes \td t + \tilde{n} \otimes \tilde{n} + \sigma +
( \dot{\tilde{n}} \otimes \tilde{n} + {\tilde{n}} \otimes \dot{\tilde{n}} ) + \dot{\sigma} \nn\\
&=&  - \tilde{c}^2 \,\td t \otimes \td t + \hat{\eta} \label{ghat}
\eea where $\dot{\tilde{n}}$ and $\dot{\sigma}$ are order of $\Or(1
/ c^2)$ and may be considered as small deformations of $S_n$ and
its normal $\tilde{n}$. We further restrict deformations of $S_n$
as follows
\be \dot{\tilde{n}} =  \frac{f}{c^2} \, {\tilde{n}},
\hspace{1.0cm} \dot{\sigma}= \frac{2\,q}{c^2} \,\sigma
\label{deformation} \ee
where $f$ and $q$ are functions of $x^\mu$. The deformation can
reduce to the weak-field limit of Schwarzschild metric in
isotropic coordinates when $f=q=GM/r$. Furthermore, it can also
yield post-Newtonian metric when $f=q=-\Phi$.  One may consider a
more general deformation, however, \eref{deformation} should be
sufficient to represent a deformation generated by a Newtonian
star.

It is convenient to introduce spherical coordinates $\{r, \theta,
\phi\}$ on $\Sigma_t$ and therefore $\hat{g}$ can be written as
\be\fl \hat{g}= - \tilde{c}^2 \,\td t \otimes \td t + (1 + \frac{2
\,f}{c^2}) \dd r \otimes \dd r + (1 + \frac{2\,q}{c^2})\, r^2 (\dd
\theta \otimes \dd \theta + \sin^2 \theta\, \dd \phi \otimes \dd
\phi ) \label{g_hat_spherical} \ee
where $f=f (t,r,\theta,\phi)$ and $q=q (t,r,\theta,\phi)$. To
solve Einstein equations
\be
 P_\a \otimes e^\a = \frac{8\, \pi G}{c^4} \left( T - \frac{1}{2} \,\hat{g} \, T( X_\a, X^\a) \right),
\ee we consider a Newtonian star which is characterized by the
stress-energy tensor $T=\rho\,c^2 \,\dd t \otimes \dd t$ and after
some tedious but straightforward calculations, the leading-order
Einstein equations yield Newtonian Poisson equation in $\{r,
\theta, \phi\}$
\be \frac{1}{r^2} \frac{\partial}{\partial r}( r^2\,
\frac{\partial \Phi}{\partial r}) + \frac{1}{r^2 \sin \theta}
\frac{\partial}{\partial \theta} (\,\sin \theta \frac{\partial
\Phi}{\partial \theta}\,) + \frac{1}{r^2 \sin^2 \theta}
\,\frac{\partial^2 \Phi}{\partial \phi^2} = 4 \pi G \rho \ee
with $f=q=- \Phi$, which agrees with post-Newtonian approximation.

In the following section, we calculate Komar, BY, and DM's
expressions in terms of \eref{g_hat_spherical} and then compare
them to \eref{Newtonian_QLM} in the limit $c \rightarrow \infty$. For simplicity, we consider $\Phi, f, q$ are only functions of $r, \theta, \phi$ in BY and DM's calculations.

\section{Quasi-local mass in the covariant Newtonian theory}
\label{Quasi}

It is well known that \eref{Newtonian_QLM} is valid for an
arbitrary closed 2-surface, it may simplify our calculations by
choosing a 2-sphere at $r=\textrm{constant}$ on $\Sigma_t$ and then
$\textbf{n}=\partial_r$. Thus, \eref{Newtonian_QLM} gives
\be M= \frac{1}{4\pi G} \oint_S\, \frac{\partial \Phi}{\partial\,
r} \,\td S = \frac{1}{4\pi G} \oint_S\, \frac{\partial
\Phi}{\partial\, r} \, r^2 \sin \theta \,\dd \theta \,\dd \phi
\label{Nmass}\ee
From \Sref{komar} to \Sref{DM}, we present  detail constructions
and calculations of the Komar integral, BY and DM's quasi-local
expressions.

\subsection{Komar integral} \label{komar}

The Komar integral is a covariant expression for quasi-local
quantities based on Killing vector fields. In terms of a time-like
Killing vector field $\xi$ with $\lim_{r\to\infty}\, \hat{g} (\xi,
\xi)=-1$, the Komar integral associated to quasi-local energy
$E_K$ can be expressed as
 \be {E_K} = -\,\frac{c^4}{8\pi  G}
\oint_S *\, \dd \tilde{\xi} \ee
where $\tilde{\xi} = \hat{g}(\xi, -)$ and $*$ denotes
4-dimensional Hodge map with $* 1= e^0 \w e^1 \w e^2 \w e^3$.
Though no time-like Killing vector field exists in spacetime with
metric $\hat{g}$, it does asymptotically approach to a time-like
Killing vector $\textbf{e}/c$ in the limit $c \to \infty$, i.e.
\be \lim_{c\to\infty} \,\frac{1}{c}\,\mathcal{L}_{\textbf{e}}\,
\hat{g}=\lim_{c\to\infty} \,\left( - \frac{2\,\partial
\Phi}{c\,\partial t} \dd t \otimes \dd t +
\Or(\frac{1}{c^3})\right) =0 \ee
where $\mathcal{L}_{\textbf{e}}$ denotes the Lie derivative with respect to $\textbf{e}$.
By using the asymptotical Killing vector field $\textbf{e}/c$,
the Komar integral yields
\bea M_K :=\lim_{c\to\infty} \,\frac{E}{c^2}&=& \lim_{c\to\infty}
-\frac{c}{8\pi G} \oint_S * \dd \,\tilde{\textbf{e}} \nn \\
&=& -\frac{1}{4\pi G} \oint_S g^\mu n_\mu \, \dd S \label{komar_integral}
\eea
where $\tilde{\textbf{e}}= \hat{g} (\textbf{e}, - )$.
\Eref{komar_integral} gives exactly the same \textit{covariant} expression as
\eref{Newtonian_QLM} without choosing a specific closed 2-surface. 

\subsection{Brown-York quasi-local energy} \label{BY}

The BY's quasi-local energy $E_{BY}$ obtained by using Hamilton-Jacobi analysis gives \cite{BY} 
\be
E_{BY} = \frac{c^4}{8 \pi G}  \oint_S ( k - k_0 )\, \dd S	 \label{BY_energy}
\ee where $k:=k^\mu{_\mu}$ is the trace of extrinsic curvature $k_{\mu\nu}$ of a closed 2-surface $S$ and $k_0$ denotes the reference values of quasi-local energy. A possible choice of $k_0$ is by   isometrically embedding $S$  into a flat 3-dimensional slice of flat spacetime as a reference space.  From \eref{g_hat_spherical}, the unit normal $\tilde{n}$ of a 2-sphere at $t, r=\textrm{constant}$ is given by
\be
\tilde{n} = (1 + \frac{f}{c^2}) \dd r
\ee and then the mean curvature $k$ yields
\bea
k = - \sigma^{\mu\nu} \gamma^\a{_\nu} (\check{\nabla}_{X_\mu}  \tilde{n} )_\a  = - \frac{2}{r} + \frac{2\,f}{c^2\, r} -  \frac{2}{c^2} \frac{\partial q}{\partial r}. \label{k}
\eea To obtain $k_0$, we start from a 3-dimensional reference space with a flat metric 
\be
\breve{\eta} = \dd R \otimes \dd R + R^2 ( \dd \theta \otimes \dd \theta + \sin^2 \theta \,\dd \phi \otimes \dd \phi  ) 
\ee By performing the coordinate transformations $r= R - r\,q / c^2$, one obtains
\bea 
\fl\breve{\eta} &=& \left(1 + \frac{2\,\partial (r q)}{ c^2\,\partial r}\right) \dd r \otimes \dd r + \frac{r\, \partial  q}{c^2\, \partial \theta} (\dd r \otimes \dd \theta + \dd \theta \otimes \dd r ) +\frac{r\, \partial  q}{c^2\, \partial \phi} (\dd r \otimes \dd \phi + \dd \phi \otimes \dd r ) \nn\\
\fl &&+\,( 1 + \frac{2\,q}{c^2} ) r^2 (\dd \theta \otimes \dd \theta + \sin^2 \theta\, \dd \varphi \otimes \dd \varphi ). 
\eea Therefore, calculating the mean curvature of $S$ at $r=\textrm{constant}$ in the reference space yields
\bea
k_0 = - \frac{2}{r} +\frac{2}{c^2 r} \frac{ \partial (r q)}{ \partial r} - \frac{2}{c^2} \frac{\partial q }{ \partial r} + \frac{1}{c^2 r}  \frac{\partial^2 q}{\partial \theta^2} + \frac{\cot\theta }{c^2 r} \frac{\partial q}{\partial \theta}+ \frac{1}{c^2 r \sin^2 \theta } \frac{\partial^2 q}{\partial \phi^2} \label{k_0}
\eea By substituting \eref{k} and \eref{k_0} into \eref{BY_energy}, we obtain
\bea 
\fl M_{BY} &:=& \lim_{c\to\infty}\frac{E_{BY}}{c^2} \nn\\
\fl &=&  \frac{1}{8 \pi G}  \oint_S \left( \frac{2\,f}{ r} - \frac{2}{ r} \frac{ \partial (r q)}{ \partial r} -\frac{1}{ r \sin \theta} \frac{\partial}{\partial \theta} (\sin \theta \frac{\partial q}{\partial \theta} ) - \frac{1}{ r \sin^2 \theta } \frac{\partial^2 q}{\partial \phi^2} \right)\dd S. \label{k-k_0}
\eea Moreover, since Einstein field equation gives $f=q=-\Phi$, $M_{BY}$ then becomes
\be
 M_{BY} =  \frac{1}{8 \pi G} \oint_S  \left( 2\, \frac{ \partial \Phi }{ \partial r} +\frac{1}{ r \sin \theta} \frac{\partial}{\partial \theta} (\sin \theta \frac{\partial \Phi}{\partial \theta} ) + \frac{1}{ r \sin^2 \theta } \frac{\partial^2 \Phi}{\partial \phi^2} \right)\dd S. \label{k-k_0-1}
\ee By comparing \eref{k-k_0-1} with \eref{Nmass}, it turns out that only when $\Phi= \Phi(r)$, i.e., spherical symmetry, \eref{k-k_0-1} can return to \eref{Nmass}. Furthermore, weak-field limit of Schwartzschild solution $\Phi=- GM / r$ yields correct result
\be
M_{BY} =  \frac{M}{4 \pi}\oint_S  \frac{1 }{ r^2}\, \dd S = M.
\ee

\subsection{Dougan-Mason quasi-local mass $(+---)$} \label{DM}

Dougan and Mason \cite{Dougan&Mason} used the holomorphic or
anti-holomorphic conditions instead of the propagation equations
of Ludvigsen-Vickers mass on a closed 2-surface, so that the Hamiltonian
depends only on the boundary values of the spinors on the
2-surface. The energy-momentum can be defined as
\bea
 P^{\underline{A}\underline{A'}}:= \frac{\cs^2}{8\pi G} \oint_S
 \mathbf{F}^{\underline{A}\underline{A'}}_{\a\b} \dd x^\a\wedge \dd x^\b = \frac{-i\cs^2}{4\pi G} \oint_S
 \mathbf{F}^{\underline{A}\underline{A'}}_{\a\b} m^\a \bar m^\b \dd S
\eea
by using Nester-Witten two-form
$\mathbf{F}^{\underline{A}\underline{A'}}=\mathbf{F}^{\underline{A}\underline{A'}}_{\a\b}
\dd x^\a \wedge \dd x^\b$ and then its components are
\bea \mathbf{F}^{\underline{A}\underline{A'}}_{\a\b}:=
-i(Z^{\underline{A}}_A \nabla_\b Z^{\underline{A'}}_{A'} -
Z^{\underline{A}}_B \nabla_\a Z^{\underline{A'}}_{B'}).\eea
The spin frame $Z^{\underline{A}}_A = (\lambda_A,\mu_A)$ which is
normalized by $ \varepsilon^{AB} \lambda_A\mu_B= 1$ such that any
solution of $\lambda_A$ can be written as $\lambda_A =
Z^{\underline{A}}_A \lambda_{\underline{A}}$ where
$\lambda_{\underline{A}}$ is constant. The indices are lowered by
the inverse $\varepsilon_{AB}$ of $\varepsilon^{AB} =
\lambda^A\mu^B-\mu^A\lambda^B$.

Generally, $\varepsilon^{AB} Z^{\underline{A}}_A
Z^{\underline{B}}_B$ may not be constant on the 2-surface $S$.
Therefore, a natural definition of
$\varepsilon^{\underline{A}\underline{B}} := \varepsilon^{AB}
Z^{\underline{A}}_A Z^{\underline{B}}_B $ demands the spinors
satisfy the holomorphic (\ref{holo}) or anti-holomorphic
(\ref{anti-holo}) equations on $S$. So given a constant
$\varepsilon^{\underline{A}\underline{B}}$, we can define the mass
by using the norm of the energy-momentum where
\bea m^2_{DM}:= P^{\underline{A}\underline{A'}}
P^{\underline{B}\underline{B'}}
\varepsilon_{\underline{A}\underline{B}}\varepsilon_{\underline{A'}\underline{B'}}
= P^{\underline{A}\underline{A'}}P_{\underline{A}\underline{A'}}.
\label{DM-mass} \eea
If we write $N= \varepsilon^{\underline{01}}$, the mass can be
calculated as (Bergqvist, \cite{Bergqvist92})
\bea m^2_{DM} = \frac{2}{|N|^2} (P^{\underline{00'}}
P^{\underline{11'}}- P^{\underline{01'}} P^{\underline{10'}}).
\label{Bergqvist-mass}\eea
Holomorphic condition is
\be\ba{lll}  \delta \lambda_A =0,  \Rightarrow  \eth \lambda_0 +
\sigma \lambda_1 =0,\hspace{1.0cm}   \eth \lambda_1 + \rho'
\lambda_0 =0. \label{holo}\ea\ee
Anti-holomorphic condition is
\be\ba{lll} \bar \delta \lambda_A =0, \Rightarrow  \eth' \lambda_1
+ \sigma' \lambda_0 =0, \hspace{1.0cm}  \eth' \lambda_0 + \rho
\lambda_1 =0.\label{anti-holo}\ea\ee
Then the quasi-local energy-momentum integral for a  2-sphere $S$
can be written as
\bea \fl I(r_S) = P^{\underline{AA'}}
\lambda_{\underline{A}}\lambda_{\underline{A'}}= \frac{\cs^2}{4\pi
G} \oint_S \{ \lambda_1 (\eth \lambda_{0'} + \rho \lambda_{1'}) -
\lambda_0(\bar\eth\lambda_{1'} + \rho' \lambda_{0'}) \} \dd S.
\eea
We use the above equation and integrate by parts, then we obtain a
very simple expression
\bea I(r_S) = \frac{C\cs^2}{4\pi G} \oint_S \{ \rho
\lambda_1\lambda_{1'} +\rho'\lambda_{0}\lambda_{0'}\} \dd S \eea
and
\bea P^{\underline{AA'}} = \frac{C\cs^2}{4\pi G} \oint_S \{ \rho
\lambda^{\underline{A}}_1\lambda^{\underline{A}}_{1'}
+\rho'\lambda^{\underline{A}}_{0}\lambda^{\underline{A}}_{0'}\}
\dd S \label{PAA}\eea
where $C=1$ for the holomorphic spinors and $C=-1$ for the
anti-holomorphic spinors. We will use the holomorphic condition in the later calculation.

We first express the four dimensional non-degenerate metric
(\ref{ghat}) by using  convention $(+---)$. Then, in terms of null frame we have
\bea \fl \hat{g}_{\mu\nu}=l_\mu n_\nu+n_\mu l_\nu- \bar m_\mu
m_\nu - m_\mu \bar m_\nu,\hspace{1.0cm}  \hat{g}^{\mu\nu}=l^\mu
n^\nu+n^\mu l^\nu- \bar m^\mu m^\nu - m^\mu \bar m^\nu \eea
where  the null tetrad can be written by ether frame and gauge
invariant $t_\mu$
\be\ba{lllllll}
            l_\mu &=& o_A o_{A'} =\cc t_\mu +
            r_\mu, & l^\mu &=& o^A o^{A'} = \frac{e^\mu}{\cc} +  r^\mu,\vspace{1 em}\\
            n_\mu &=&\iota_A \iota_{A'}=\half(\cc t_\mu - r_\mu), & n^\mu &=& \iota^A \iota^{A'}
            =\frac{e^\mu}{2 \cc} -\frac{r^\mu}{2},\vspace{1 em}\\
             m_\mu &=&o_A \iota_{A'} = \frac{1}{\sqrt{2}}(x_\mu - i y_\mu), & m^\mu &=& o^A \iota^{A'}
              = \frac{1}{\sqrt{2}}(x^\mu - i y^\mu).\vspace{1 em}
\label{lnm} \ea\ee
The null frame satisfied the orthogonal relations $l^\mu n_\mu =1,
m^\mu \bar m_\mu=-1$. Also, the ether frame $e^\mu$ and $t_\mu$
can be expressed in terms of null frame where
\bea t_\mu &=& \frac{1}{\cc} (\frac{l_\mu}{2}+
n_\mu),\hspace{2.0cm}
      r_\mu = \frac{l_\mu}{2} - n_\mu ,\\
      e^\mu &=& \cc (\frac{l^\mu}{2} + n^\mu),\hspace{2.0cm}
       r^\mu=  \frac{l^\mu}{2} - n^\mu,\label{ter}
      \eea
and it satisfied the orthogonal relations $e^\mu t_\mu =1, r^\mu
r_\mu = -1$ in the covariant Newtonian theory. Thus we chose
$e^\mu=(1,0,0,0)$ and $t_\mu=(1,0,0,0)$ in $(t,r,\theta,\phi)$
coordinate and use the metric
\bea \dd s^2 = \cc^2 \dd t^2 - (1+\frac{f}{c^2})^2 \dd r^2
-(r+\frac{g}{c^2})^2 \dd \theta^2 -(r+\frac{g}{c^2})^2\sin^2\theta
\dd \phi^2\eea
with leading-order correction term $f(r,\theta,\phi)$ and
$g(r,\theta,\phi)$. By choosing a corresponding null frame
according to (\ref{lnm})-(\ref{ter}), we have
\be\ba{llllll} l_\mu &=&[\cc, -(1 +\frac{f}{c^2}), 0, 0] & l^\mu &=& [\cc^{-1},(1 +\frac{f}{c^2})^{-1}, 0,0],\nn \vspace{1 em} \\
     n_\mu &=&[\half \cc, \half(1 +\frac{f}{c^2}), 0, 0]
             & n^\mu &=& \half[\cc^{-1},-(1 +\frac{f}{c^2})^{-1}, 0,0 ]\nn \vspace{1 em}\\
     m_\mu &=& [0, 0, -\frac{r'}{\sqrt{2}}, i \frac{r'}{\sqrt{2}} \sin\theta ]
     & m^\mu &=& [0,0, \frac{1}{\sqrt{2} r' },\frac{- i}{\sqrt{2} r' \sin\theta}]\vspace{1 em}
      \nn \ea\ee
where $r':=r+g/c^2$. The NP spin coefficients have the falloff
\bea  \fl \kappa=\nu=\tau=\pi=\epsilon=\c = \Or(\frac{1}{c^2}),
\hspace{1.5cm} \sigma=\lambda=0,\hspace{1.5cm}
\a=\b=\rho=\mu=\Or(1), \nn\eea
with respect to $c$.

The directional derivative along $m$ can be expanded as
\bea \d = m^\mu \nabla_\mu=\up{0}\d + \frac{\up{2}\d}{c^2} +
\Or(\frac{1}{c^4})\eea
where
\bea \up{0}\d := \frac{1}{\sqrt{2}
r}(\partial_\theta-\frac{i}{\sin\theta}
\partial_\phi), \hspace{2.0cm} \up{2}\d := - \frac{g}{r} \up{0}\d
. \eea

The key NP coefficients that we will use later are $\mu, \rho, \b$
and after some calculations yield
\bea\fl
     \rho = -\frac{1}{r} -\frac{\up{2}\rho}{rc^2}+ \Or(\frac{1}{c^4}),\hspace{1.0cm}
     \mu = -\frac{1}{2 r} - \frac{\up{2}\mu}{ r c^2} + \Or(\frac{1}{c^4}), \hspace{1.0cm}
     \b =  \up{0}\b + \frac{\up{2}\b}{c^2} + \Or(\frac{1}{c^4}),\nn\eea
where
\bea \fl \up{2}\mu =\frac{\up{2}\rho}{2},\hspace{1.0cm}
\up{2}\rho:=
\partial_r g -f - \frac{g}{r},\hspace{1.0cm}  \up{0}\b:=\frac{\sqrt{2}
\cot\theta}{4 r},\hspace{2.0cm}
 \up{2}\b:= -\frac{g}{r}\up{0}\b + \frac{\up{0}\d g}{2r}. \eea
 The asymptotic constant spinor $\lambda_A = \lambda_1 o_A -\lambda_0 \iota_A$ can be expressed as
\bea \lambda_0 = \up{0}\lambda_0  + \frac{\up{2}\lambda_0}{c^2} +
\Or(\frac{1}{\cs^4}),\hspace{1.0cm}  \lambda_1 = \up{0}\lambda_1 +
\frac{\up{2}\lambda_1}{\cs^2} + \Or(\frac{1}{\cs^4}). \eea

By substituting the relevant asymptotic expansion terms into the
surface integration, we have
 \bea 
 \fl & &\rho \lambda_1 \lambda_{1'} - \mu \lambda_0 \lambda_{0'} \nn \vspace{1 em} \\
     \fl &=&\up{0}\rho \up{0}\lambda_1 \up{0}\lambda_{1'}- \up{0}\mu \up{0}\lambda_0 \up{0}\lambda_{0'}\vspace{1 em}\label{surface1}\\
    \fl & &+ \frac{1}{c^2}[\up{2}\rho \up{0}\lambda_1 \up{0}\lambda_{1'}- \up{2}\mu \up{0}\lambda_0 \up{0}\lambda_{0'}
      + \up{0}\rho \up{2}\lambda_1 \up{0}\lambda_{1'}- \up{0}\mu \up{2}\lambda_0 \up{0}\lambda_{0'}
              +\up{0}\rho \up{0}\lambda_1 \up{2}\lambda_{1'}- \up{0}\mu \up{0}\lambda_0
              \up{2}\lambda_{0'}] + \Or(\frac{1}{c^4}). \label{surface2}
            \eea
We will show that only (\ref{surface2}) will survive.

The first order $\Or(1)$ holomorphic conditions are
\bea \up{0}\d \up{0}\lambda_0 - \up{0}\b \up{0}\lambda_0
=0,\hspace{2.0cm}
     \up{0}\d \up{0}\lambda_1 + \up{0}\b \up{0} \lambda_1 - \up{0}\mu \up{0}\lambda_0 =0.\eea
Therefore, the solutions are
\bea \fl\up{0} \lambda_0 = C_{-1} \up{0}A_{-1}
\exp(-i\frac{\phi}{2}) + C_1 \up{0}A_1
\exp(i\frac{\phi}{2}),\hspace{1.0cm} \up{0} \lambda_1 =
\up{0}B_{-1} \exp(-i\frac{\phi}{2}) + \up{0}B_1
\exp(i\frac{\phi}{2})\eea
where
 \bea \up{0}A_{-1} &=& \sin\frac{\theta}{2},
\hspace{3.5cm} \up{0}A_{1}
=\cos\frac{\theta}{2},\\
\up{0}B_{-1} &=& C_1 \frac{\cos\frac{\theta}{2}}{\sqrt{2}} =C_1
\frac{\up{0}A_1}{\sqrt{2}}, \hspace{1.5cm} 
\up{0}B_{1} = -C_{-1}\frac{\sin\frac{\theta}{2}}{\sqrt{2}} =
-C_{-1}\frac{\up{0}A_{-1}}{\sqrt{2}}.
 \eea
 We substitute the first order spinors and the related NP
coefficients into (\ref{surface1}) and integrate it.  We then find $\up{0}{P^{00'}}$ vanish.
%
%
Furthermore, we can find $\up{0}{P^{00'}}=\up{0}{P^{11'}}$ from
symmetry of the solution and $\up{0}{P^{10'}}=\up{0}{P^{01'}}=0$
from $\int^{2\pi}_0 \exp(i \phi) \dd \phi =0$. The second order
$\Or(1/c^2)$ holomorphic differential equations are:
\bea (\up{0}\d  -\up{0}\b)\up{2}\lambda_0 - \frac{g}{2r} \up{0}\lambda_0 =0,\\
     (\up{0}\d  + \up{0}\b) \up{2}\lambda_1 - \up{0}\mu \up{2} \lambda_0 +\frac{g}{2r} \up{0}\lambda_1
     +\frac{\partial_r g - f}{2r} \up{0}\lambda_0 =0,\eea
where
\bea \fl \up{2} \lambda_0 = \up{2}A_{-1} \exp(-i\frac{\phi}{2}) +
\up{2}A_1 \exp(i\frac{\phi}{2}), \hspace{1.0cm}  \up{2} \lambda_1
= \up{2}B_{-1} \exp(-i\frac{\phi}{2}) + \up{2}B_1
\exp(i\frac{\phi}{2}),\eea
and
\bea  \up{2}A_{-1} &=& C_{-1}\int^{\pi}_{\theta} \up{0}\d g \dd
\theta' \up{0}A_{-1},\hspace{2.0cm}  \up{2}A_{1}=
C_{1}\int^{\theta}_{0} \up{0}\d g \dd \theta' \up{0}A_{1} , \eea
\bea \fl \up{2}B_{-1}&=& -
\frac{1}{\sqrt{2}\up{0}A_1}\int^{\pi}_{\theta}[ C_{-1}
\frac{\int^\pi_\theta \up{0}\d g \dd \theta'
\up{0}A_{-1}}{\sqrt{2}} + \up{0}\d g \up{0}B_{-1}
+ C_{-1}(\partial_r g -f)\up{0}A_{-1}]\up{0}A_{1} \dd \theta, \\
\fl \up{2}B_{1} &=& - \frac{1}{\sqrt{2}
\up{0}A_{-1}}\int^{\theta}_{0}[ C_{1} \frac{\int^{\theta}_0
\up{0}\d g \dd \theta' \up{0}A_1}{\sqrt{2}} + \up{0}\d g
\up{0}B_{1} + C_{1}(\partial_r g -f)\up{0}A_{1}]\up{0}A_{-1} \dd
\theta , \eea
where $\up{2}A_{-1},\up{2}A_{1},\up{2}B_{-1},\up{2}B_{1}$ at
$0,\pi$ vanish.

We can further find $\oint
\up{0}{\lambda_1^0}\up{2}{\lambda_{1'}^{1'}} \dd S = \oint
\up{0}{\lambda_1^1}\up{2}{\lambda_{1'}^{0'}} \dd S =0 $ and
$\up{0}{\lambda_1^1}\up{2}{\lambda_{1'}^{1'}}=
\up{0}{\lambda_1^0}\up{2}{\lambda_{1'}^{0'}}$. Therefore, only
$(00')$ component will contribute to the energy-momentum surface
integration. Thus we substitute the second order spinors and NP
coefficients into the $(00')$ component of (\ref{surface2}), we
then find
\bea \fl \up{2}{P^{00'}} =\frac{r}{8G}\int^\pi_0 (\partial_r g
-f)\, \sin(2\theta)\, \dd \theta -\frac{r}{4 G} \int^\pi_0
\int_{\theta'}^\pi (\partial_r g -f) \,\sin\theta \, \sin\theta'\,
\dd \theta \, \dd \theta' \nn\eea
and
\bea |N| &=& |{\lambda^0_0}{\lambda^1_1}-{\lambda^0_1}
{\lambda^1_0}| =\frac{1}{\sqrt{2}} + \Or(\frac{1}{c^2}).\eea
The mass expression (\ref{DM-mass}) can be expanded as
\bea \fl \lim_{c\to\infty} m_{DM} &=& \frac{\sqrt{2} \up{2}{P^{00'}}}{|\up{0}N|} \\
\fl &=& \frac{r}{4G}\int^\pi_0 (\partial_r g -f)\, \sin(2\theta) \, \dd \theta 
 -\frac{r}{2 G} \int^\pi_0 \int_{\theta'}^\pi
(\partial_r g -f) \,\sin\theta \, \sin\theta'\, \dd \theta \dd \,
\theta'  \nn\eea
For Einstein field equation satisfied, we have
$f=g/r=-\Phi(r,\theta,\phi)$. If we consider spherical symmetry
of $g, f$ and $f = G M/r = -\Phi(r)$ and $g= -r \Phi(r)$, then the DM mass becomes
the quasi-local Newtonian mass
\bea  \lim_{c\to\infty} m_{DM}  
&=& \frac{1}{4\pi G} \oint_S \partial_r \Phi(r)\, \dd S. \eea

\section{Conclusion}

In this work, we verify that Komar integral can yield the Newtonian
quasi-local mass expression without choosing a specific 2-sphere or referring to spherical symmetry of the Newtonian potential $\Phi$, however, the Brown-York expression and
Dougan-Mason mass can give Newtonian expression only in the spherical symmetry of $\Phi$. Since general relativity has a well known Newtonian limit, it is reasonable for us to argue that it should yield the Newtonian expression for all of quasi-local expressions in the covariant Newtonian space-time. We discover that the differences between Newtonian expression and $M_{BY}$ are contributed from reference of energy surface density $k_0$ which is obtained by isometrically embedding a 2-sphere in a flat reference 3-space. Moreover, we find that the Newtonian source will distort not only NP spin coefficients but also the spin frame $\lambda_A$ for the
leading-order term of the Dougan-Mason mass expression.


The Newtonian theory is sufficient enough to describe the most of
the astrophysical problems. It would be an interesting question
that can we measure the correct amount of mass for an arbitrary
Newtonian source by using these quasi-local expressions?
Unfortunately, we find that only for the spherical symmetry
source Brown-York and Dougan-Mason mass can get the surface
integration of the Newtonian mass. Whether this problem is due to
the expressions themselves or some other technical problems requires a further
investigation.


\ack YHW would like to thank
her host Prof Chung-Ming Ko and she is financially supported by
the National Science Council (NSC) of ROC, Grant No. NSC 96-2811-M-008-056. CHW thanks Mr. Jian-Liang Liu for helpful discussions. CHW is supported by NSC under Grant No.
NSC 096-2811-M-008-040.

\end{document}